# Successive Cancellation List Decoding of BMERA Codes with Application to Higher-Order Modulation

Tobias Prinz, Peihong Yuan
Institute for Communications Engineering
Technical University of Munich
Email: tobias.prinz@tum.de, peihong.yuan@tum.de

*Abstract*—BMERA or convolutional polar codes are an extension of polar codes with a provably better error exponent than polar codes. A successive cancellation (SC) decoding algorithm for BMERA codes similar to SC polar decoders is introduced. A pseudocode description of the SC decoder that can be extended to SC list (SCL) decoding is provided. Simulation results with and without outer CRC codes under SC and SCL decoding are presented for QAM modulation over the AWGN channel to compare the performance of polar and BMERA codes. BMERA codes outperform polar codes by more than 0.5 dB under SCL decoding without outer CRC codes.

## I. Introduction

Polar codes were proposed in [1], [2] and it was proven in [1] that they achieve the capacity of binary input discrete memoryless channels asymptotically in block length with an error exponent of $1/2$ under successive cancellation (SC) decoding. To increase the error exponent, polar codes with larger or non-binary kernels can be used [3].

Branching multi-scale entanglement renormalization ansatz (BMERA) codes are introduced in [4] as a natural extension of polar codes. An error exponent of $0.5 \log_2 3$ is proven for BMERA codes under SC decoding [5], but no description for the SC decoder is given. Besides, simulation results for SC decoding of BMERA codes are provided using graphical calculus tools based on tensor networks [4]. A first explicit SC decoder for BMERA codes is introduced in [6], [7].

The recursive formulas for SC decoding of BMERA codes and results for SC list (SCL) decoding are presented in [8]. The main contribution of this paper is to provide a pseudocode description of SC decoding for BMERA codes with a similar structure as for polar codes in [9] and simulation results for higher-order modulation schemes (QAM).

The paper is structured as follows. In Sec. II, we introduce the BMERA transform and relate it to the polar one. In Sec. III, the recursive formulas for BMERA decoding are derived and a pseudocode description of the decoding algorithm is presented. Sec. IV briefly summarizes how BMERA codes can be used for higher-order modulation schemes. Numerical simulation results for BPSK and QAM modulation are presented in Sec. V.

## II. Polar and BMERA transform

The polar transform [1] for block length $n$ is defined as $\mathbf{c} = \mathbf{u} \boldsymbol{B}_n \mathbb{F}^{\otimes \log_2 n}$, where $\mathbf{u}, \mathbf{c}$ are binary vectors of length $n$,

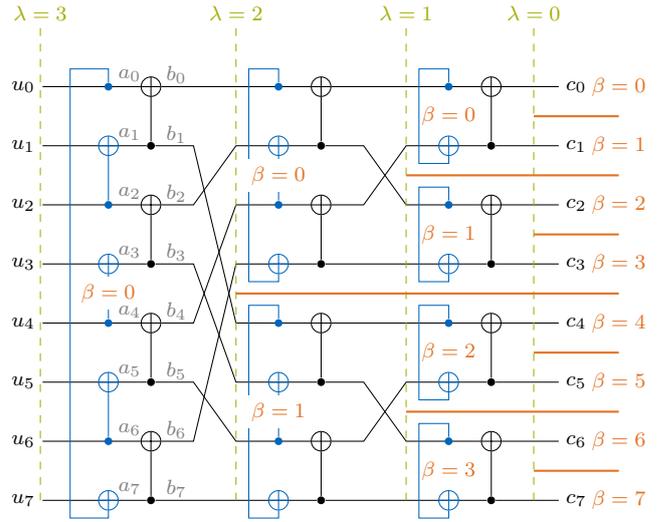

Fig. 1. Graph representation of the BMERA transform for $n = 8$.

$\boldsymbol{B}_n$ is the bit reversal matrix of size $n$ and the polarization kernel is

$$\mathbb{F} = \begin{bmatrix} 1 & 0 \\ 1 & 1 \end{bmatrix}. \quad (1)$$

The graph in Fig. 1 without the blue connections illustrates a polar transform of size $n = 8$ in bit-reversal representation.

By adding the blue lines in Fig. 1, we obtain a BMERA transform [4]. The BMERA transform of size $n = 2^m$ without bit-reversal permutations is recursively defined as

$$\boldsymbol{G}_n = \left[ \boldsymbol{S}_n \left( \boldsymbol{I}_{n/2} \otimes \mathbb{F} \right) \boldsymbol{S}_n^{\mathrm{T}} \left( \boldsymbol{I}_{n/2} \otimes \mathbb{F} \right) \right] \cdot \left( \boldsymbol{G}_{n/2} \otimes \boldsymbol{I}_2 \right) \quad (2)$$

where $\boldsymbol{G}_1 = 1$ and $\boldsymbol{S}_n$ is the identity matrix of size $n$ whose columns are circularly shifted to the left by one. The BMERA transform with bit-reversal permutation as depicted in Fig. 1 is then given as

$$\mathbf{c} = \mathbf{u} \boldsymbol{B}_n \boldsymbol{G}_n. \quad (3)$$

For BMERA and polar codes $\mathbf{c}$ is modulated to the channel input symbols $\mathbf{x}$, e.g., $0 \to +1$ and $1 \to -1$ for binary phase shift keying (BPSK). The received signal is denoted by $\mathbf{y}$.

A different BMERA transform is used in [8], where the circular connections in Fig. 1 are omitted. In [5], this is called BMERA with open boundaries and the transform used in this paper is called BMERA with closed boundaries.

## III. BMERA Decoding

We use a similar notation as in [9] for the recursive description of polar decoding. The notation is illustrated in Fig. 1. For a code of length $n = 2^m$, we have $m + 1$ *layers* indexed by $\lambda$, where $0 \leq \lambda \leq m$. Within a layer $\lambda$, we have $2^{m-\lambda}$ *branches* indexed by $\beta$, where $0 \leq \beta < 2^{m-\lambda}$. The bits within the branch of a layer are indexed by the *phase* $\varphi$, where $0 \leq \varphi < 2^\lambda = \Lambda$.

SC decoding for BMERA and polar codes successively decodes the bits $\mathsf{u}_\varphi$ based on the channel outputs $\mathbf{y}$ and the previously decoded bits $\mathbf{u}_0^{\varphi-1}$, i.e., we evaluate

$$P_{U_\varphi | Y_0^{n-1} U_0^{\varphi-1}} \left( \mathsf{u}_\varphi | \mathbf{y}, \mathbf{u}_0^{\varphi-1} \right) \quad (4)$$

for $\mathsf{u}_\varphi = \{0, 1\}$. For polar codes, the relations between $U_\varphi$ and $\left( Y_0^{\varphi-1}, U_0^{\varphi-1} \right)$ are referred to as *bit-channels* and (4) can be recursively calculated from the bit-channels of previous layers. In the following, we derive the recursive formulas for BMERA codes.

### A. 3-Bit Channels for BMERA Codes

Using the notation of $a_\varphi$'s and $b_\varphi$'s illustrated in Fig. 1, we can express (4) for BMERA codes for odd $\varphi < n - 1$ as

$$\begin{aligned} P_{U_\varphi | Y_0^{n-1} U_0^{\varphi-1}} &\stackrel{(a)}{=} \sum_{u_{\varphi+1}} P_{U_\varphi U_{\varphi+1} | Y_0^{n-1} U_0^{\varphi-1}} \\ &\stackrel{(b)}{=} \sum_{u_\varphi+1} P_{A_\varphi A_{\varphi+1} | Y_0^{n-1} A_0^{\varphi-1}} \\ &\stackrel{(c)}{=} \frac{1}{P_{A_{\varphi-1} | Y_0^{n-1} A_0^{\varphi-2}}} \sum_{u_\varphi+1} P_{A_{\varphi-1} A_\varphi A_{\varphi+1} | Y_0^{n-1} A_0^{\varphi-2}} \\ &\stackrel{(d)}{=} \frac{1}{c} \sum_{a_{\varphi+2}} \sum_{u_{\varphi+1}} P_{A_{\varphi-1} A_\varphi A_{\varphi+1} A_{\varphi+2} | Y_0^{n-1} A_0^{\varphi-2}} \\ &\stackrel{(e)}{=} \frac{1}{c} \sum_{a_{\varphi+2}} \sum_{u_{\varphi+1}} P_{B_{\varphi-1} B_\varphi B_{\varphi+1} B_{\varphi+2} | Y_0^{n-1} B_0^{\varphi-2}} \\ &\stackrel{(f)}{=} \frac{1}{c} \sum_{a_{\varphi+2}} \sum_{u_{\varphi+1}} P_{B_{\varphi-1} B_{\varphi+1} | Y_0^{n/2-1} B_0 B_3 \cdots B_{\varphi-3}} \\ &\qquad \cdot P_{B_\varphi B_{\varphi+2} | Y_{n/2}^{n-1} B_1 B_2 \cdots B_{\varphi-2}} \end{aligned} \quad (5)$$

where $(a)$ and $(d)$ follow from marginalization, $(b)$ follows from the one-to-one relation between $U_\varphi, U_{\varphi+1}$ and $A_\varphi, A_{\varphi+1}$ and between $U_0^{\varphi-1}$ and $A_0^{\varphi-1}$, $(c)$ follows from Bayes' rule, $(e)$ follows from the one-to-one relations between the corresponding $A_i$'s and $B_i$'s and $(f)$ from the independence of the two sub-branches.

Equation (5) shows that it is not possible to calculate (4) recursively from the bit-channels of the previous layer. Similarly, the joint distribution $P_{U_\varphi U_{\varphi+1} | Y_0^{n-1} U_0^{\varphi-1}}$ can not be calculated from the joint distributions of two bits of the previous layer for all $0 \leq \varphi \leq n - 2$.

Considering *3-bit channels* [5], we can compute the distributions $P_{U_\varphi U_{\varphi+1} U_{\varphi+2} | Y_0^{n-1} U_0^{\varphi-1}}$ for all $0 \leq \varphi \leq n-3$ recursively from the 3-bit-channels of the previous layer. In the following we will state the recursive formulas for all $0 \leq \varphi \leq n - 3$, where we consider four cases for different values of $\varphi$.

### B. 3-Bit-Channel Distributions

In order to understand the following formulas with the help of Fig.1, we neglect the arguments of the distributions and use the notation of the auxiliary variables $A_i$ and $B_i$. To determine the following formulas, we can again use marginalization, one-to-one relations, Bayes' rule and the independence of branches in a similar way as in (5).

For odd $\varphi$, $\varphi < n - 3$, we obtain

$$\begin{aligned} &P_{U_\varphi^{\varphi+2} | Y_0^{n-1} U_0^{\varphi-1}} \\ &= \frac{1}{c} \sum_{u_{\varphi+3}} \sum_{a_{\varphi+4}} P_{B_{\varphi-1} B_{\varphi+1} B_{\varphi+3} | Y_0^{n/2-1} B_0 B_2 \cdots B_{\varphi-3}} \\ &\qquad \cdot P_{B_\varphi B_{\varphi+2} B_{\varphi+4} | Y_n/2^{n-1} B_1 B_3 \cdots B_{\varphi-2}}. \end{aligned} \quad (6)$$

For odd $\varphi = n - 3$ (the last three bits in a layer), the two marginalizations in (6) drop out and we have

$$\begin{aligned} P_{U_\varphi^{\varphi+2} | Y_0^{n-1} U_0^{\varphi-1}} &= \frac{1}{c} P_{B_{\varphi-1} B_{\varphi+1} | Y_0^{n/2-1} B_0 B_2 \cdots B_{\varphi-3}} \\ &\quad \cdot P_{B_\varphi B_{\varphi+2} | Y_n/2^{n-1} B_1 B_3 \cdots B_{\varphi-2}}. \end{aligned} \quad (7)$$

For (7) we need only 2-bit channels of the previous layer. In order to get a consistent representation with 3-bit channels, we can use Bayes' rule and obtain

$$\begin{aligned} P_{U_\varphi^{\varphi+2} | Y_0^{n-1} U_0^{\varphi-1}} &= \frac{1}{c'} P_{B_{\varphi-3} B_{\varphi-1} B_{\varphi+1} | Y_0^{n/2-1} B_0 B_2 \cdots B_{\varphi-5}} \\ &\quad \cdot P_{B_{\varphi-2} B_\varphi B_{\varphi+2} | Y_n/2^{n-1} B_1 B_3 \cdots B_{\varphi-4}}. \end{aligned} \quad (8)$$

For even $\varphi$, $0 < \varphi \leq n - 4$, we obtain

$$\begin{aligned} P_{U_\varphi^{\varphi+2} | Y_0^{n-1} U_0^{\varphi-1}} &= \\ &= \frac{1}{c} \sum_{a_{\varphi+3}} P_{B_{\varphi-2} B_\varphi B_{\varphi+2} | Y_0^{n/2-1} B_0 B_2 \cdots B_{\varphi-4}} \\ &\quad \cdot P_{B_{\varphi-1} B_{\varphi+1} B_{\varphi+3} | Y_{n/2}^{n-1} B_1 B_3 \cdots B_{\varphi-3}}. \end{aligned} \quad (9)$$

For $\varphi = 0$, the two applications of Bayes' rule in (9) drop out and we have

$$P_{U_\varphi^{\varphi+2} | Y_0^{n-1}} = \sum_{a_{\varphi+3}} P_{B_\varphi B_{\varphi+2} | Y_0^{n/2-1}} P_{B_{\varphi+1} B_{\varphi+3} | Y_{n/2}^{n-1}}. \quad (10)$$

Marginalizing over $B_{\lambda+4}$ and $B_{\lambda+5}$, we get the 3-bit-channel representation

$$P_{U_\varphi^{\varphi+2} | Y_0^{n-1}} = \sum_{a_{\varphi+3}} \sum_{b_{\varphi+4}} \sum_{b_{\varphi+5}} P_{B_\varphi B_{\varphi+2} B_{\varphi+4} | Y_0^{n/2-1}} \\ \cdot P_{B_{\varphi+1} B_{\varphi+3} B_{\varphi+5} f | Y_{n/2}^{n-1}}. \quad (11)$$

### C. Recursive Formulations

Having derived the recursive formulations in (6)–(11), we now omit the auxiliary variables $A_i$ and $B_i$ and resort to the notation in [9]. Let

$$W_{\lambda,\beta}^{(\varphi)}(u_\varphi, u_{\varphi+1}, u_{\varphi+2} \mid \boldsymbol{y}_0^{\Lambda-1}, \boldsymbol{u}_0^{\varphi-1}) \quad (12)$$

be the joint a posteriori probability of bits $\varphi, \varphi + 1, \varphi + 2$ in branch $\beta$ of layer $\lambda$ given all the previous bits in this branch. All channel outputs that are associated with this branch are denoted by $\mathbf{y}_0^{\Lambda-1}$ where $\mathbf{y}_0^{\Lambda-1}$ is a subvector of $\mathbf{y}$.[1] Let $\psi = \lfloor (\varphi-1)/2 \rfloor$. We use the variables $t_0, t_1, \ldots \in \{0,1\}$ for marginalizations and write multiple sums as $\sum_{t_0,t_1,\ldots} \triangleq \sum_{t_0=0}^{1} \sum_{t_1=0}^{1} \cdots$.

We can express (6) for odd $\varphi$, $\varphi < 2^\lambda - 3$ by

$$W_{\lambda,\beta}^{(\varphi)}(u_\varphi, u_{\varphi+1}, u_{\varphi+2} \mid \mathbf{y}_0^{\Lambda-1} \mathbf{u}_0^{\varphi-1})$$
$$= \frac{1}{c} \sum_{t_0,t_1} W_{\lambda-1,2\beta}^{(\psi)}(u_{\varphi-1} \oplus u_\varphi \oplus u_{\varphi+1}, u_{\varphi+1} \oplus u_{\varphi+2} \oplus t_0,$$
$$t_0 \oplus t_1 \mid \mathbf{y}_0^{\Lambda/2-1}, \mathbf{u}_{0,\text{even}}^{\varphi-3} \oplus \mathbf{u}_{1,\text{odd}}^{\varphi-2} \oplus \mathbf{u}_{2,\text{even}}^{\varphi-1})$$
$$\cdot W_{\lambda-1,2\beta+1}^{(\psi)}(u_\varphi \oplus u_{\varphi+1}, u_{\varphi+2} \oplus t_0, t_1 \mid$$
$$\mathbf{y}_{\Lambda/2}^{\Lambda-1}, \mathbf{u}_{1,\text{odd}}^{\varphi-2} \oplus \mathbf{u}_{2,\text{even}}^{\varphi-1}). \quad (13)$$

For $\varphi = 2^\lambda - 3$ we can express (8) by

$$W_{\lambda,\beta}^{(\varphi)}(u_\varphi, u_{\varphi+1}, u_{\varphi+2} \mid \mathbf{y}_0^{\Lambda-1} \mathbf{u}_0^{\varphi-1})$$
$$= \frac{1}{c} W_{\lambda-1,2\beta}^{(\psi-1)}(u_{\varphi-3} \oplus u_{\varphi-2} \oplus u_{\varphi-1}, u_{\varphi-1} \oplus u_\varphi \oplus u_{\varphi+1},$$
$$u_{\varphi+1} \oplus u_{\varphi+2} \oplus u_0 \mid \mathbf{y}_0^{\Lambda/2-1}, \mathbf{u}_{0,\text{even}}^{\varphi-5} \oplus \mathbf{u}_{1,\text{odd}}^{\varphi-4} \oplus \mathbf{u}_{2,\text{even}}^{\varphi-3})$$
$$\cdot W_{\lambda-1,2\beta+1}^{(\psi-1)}(u_{\varphi-2} \oplus u_{\varphi-1}, u_\varphi \oplus u_{\varphi+1}, u_{\varphi+2} \oplus u_0 \mid$$
$$\mathbf{y}_{\Lambda/2}^{\Lambda-1}, \mathbf{u}_{1,\text{odd}}^{\varphi-4} \oplus \mathbf{u}_{2,\text{even}}^{\varphi-3}). \quad (14)$$

For even $\varphi$, $0 < \varphi \leq n-4$, we write (9) recursively as

$$W_{\lambda,\beta}^{(\varphi)}(u_\varphi, u_{\varphi+1}, u_{\varphi+2} \mid \mathbf{y}_0^{\Lambda-1} \mathbf{u}_0^{\varphi-1})$$
$$= \frac{1}{c} \sum_{t_0} W_{\lambda-1,2\beta}^{(\psi)}(u_{\varphi-2} \oplus u_{\varphi-1} \oplus u_\varphi, u_\varphi \oplus u_{\varphi+1} \oplus u_{\varphi+2},$$
$$u_{\varphi+2} \oplus t_0 \mid \mathbf{y}_0^{\Lambda/2-1}, \mathbf{u}_{0,\text{even}}^{\varphi-4} \oplus \mathbf{u}_{1,\text{odd}}^{\varphi-3} \oplus \mathbf{u}_{2,\text{even}}^{\varphi-2})$$
$$\cdot W_{\lambda-1,2\beta+1}^{(\psi)}(u_{\varphi-1} \oplus u_\varphi, u_{\varphi+1} \oplus u_{\varphi+2}, t_0 \mid$$
$$\mathbf{y}_0^{\Lambda/2-1}, \mathbf{u}_{1,\text{odd}}^{\varphi-3} \oplus \mathbf{u}_{2,\text{even}}^{\varphi-2}). \quad (15)$$

For $\varphi = 0$ we write (8) as

$$W_{\lambda,\beta}^{(\varphi)}(u_\varphi, u_{\varphi+1}, u_{\varphi+2} \mid \mathbf{y}_0^{\Lambda-1} \mathbf{u}_0^{\varphi-1})$$
$$= \frac{1}{c} \sum_{t_0,t_1,t_2} W_{\lambda-1,2\beta}^{(\psi+1)}(u_0 \oplus u_1 \oplus u_2, u_2 \oplus t_0, t_1 \mid \mathbf{y}_0^{\Lambda/2-1})$$
$$\cdot W_{\lambda-1,2\beta+1}^{(\psi+1)}(u_1 \oplus u_2, t_0, t_2 \mid \mathbf{y}_{\Lambda/2}^{\Lambda-1}). \quad (16)$$

The normalization term $c$ in (13)–(16) must be chosen such that

$$\sum_{u_\varphi, u_{\varphi+1}, u_{\varphi+2}} W_{\lambda,\beta}^{(\varphi)}\left(u_\varphi, u_{\varphi+1}, u_{\varphi+2} \mid \mathbf{y}_0^{\Lambda-1}, \mathbf{u}_0^{\varphi-1}\right) = 1.$$

The stopping criteria for the recursion is reached in layer $\lambda = 2$. This is the first layer with more than three bits per branch and we can initialize the 3-bit probabilities. We have

$$W_{2,\beta}^{(0)}(u_0, u_1, u_2 \mid \mathbf{y}_0^{\Lambda-1}) = \sum_{u_3} \prod_{i=0}^{3} W_{0,4\beta+i}^{(0)}(c_i \mid y_i) \quad (17)$$

[1] Note that we use sans-serif letters for the whole codeword $\mathbf{c}$ and the corresponding $\mathbf{u}$ from (3) and channel output $\mathbf{y}$. By $\mathbf{u}$ we denote the bits of a certain layer $\lambda$ and branch $\beta$ and by $\mathbf{y}$ we denote the channel outputs corresponding to a branch $\beta$, where $\lambda$ and $\beta$ follow from the context.

---

**Algorithm 1:** SC Decoder

**Input:** $P_{X_i|Y_i}(x_i|\mathsf{y}_i) \forall 0 \leq i \leq n-1$ and the frozen bits
**Output:** $\hat{\mathbf{u}}$ and $\hat{\mathbf{c}}$

1 **for** $\varphi = 0, 1, \ldots, n-1$ **do**
2    $P_{0,\varphi}[0] \leftarrow P_{X_\varphi|Y_\varphi}(0|\mathsf{y}_\varphi)$
3    $P_{0,\varphi}[1] \leftarrow P_{X_\varphi|Y_\varphi}(1|\mathsf{y}_\varphi)$
4 **for** $\varphi = 0, 1, \ldots, n-3$ **do**                    // main loop
5    recursivelyCalcP$(m, \varphi)$
6    **if** $\mathsf{u}_\varphi$ *is frozen* **then**
7      $B_{m,0}[\varphi] = 0$
8    **else**
9      **if** $\sum_{t_0,t_1} P_m[0][(0,t_0,t_1)] > 0.5$ **then**
10        $B_{m,0}[\varphi] = 0$
11      **else**
12        $B_{m,0}[\varphi] = 1$
13    recursivelyUpdateB$(m, \varphi)$
14 **if** $\mathsf{u}_{n-2}$ *is frozen* **then**
15    $B_{m,0}[n-2] = 0$
16 **else**
17    $p_0 = \sum_{t_0} P_{m,0}[B_{m,0}[n-3], 0, t_0]$
18    $p_1 = \sum_{t_0} P_{m,0}[B_{m,0}[n-3], 1, t_0]$
19    **if** $p_0 > p_1$ **then**
20      $B_{m,0}[n-2] = 0$
21    **else**
22      $B_{m,0}[n-2] = 1$
23 **if** $\mathsf{u}_{n-1}$ *is frozen* **then**
24    $B_{m,0}[n-1] = 0$
25 **else**
26    $p_0 = \sum_{t_0} P_{m,0}[B_{m,0}[n-3], B_{m,0}[n-2], 0]$
27    $p_1 = \sum_{t_0} P_{m,0}[B_{m,0}[n-3], B_{m,0}[n-2], 1]$
28    **if** $p_0 > p_1$ **then**
29      $B_{m,0}[n-1] = 0$
30    **else**
31      $B_{m,0}[n-1] = 1$
32 recursivelyUpdateB$(m, n-1)$
33 $\hat{\mathbf{u}} \leftarrow B_{m,0}[0, 1, \ldots, n-1]$
34 $\hat{\mathbf{c}} \leftarrow (B_{m,0}[0], B_{m,1}[0], \ldots, B_{m,n-1}[0])$

---

for $\varphi = 0$ and

$$W_{2,\beta}^{(1)}(u_1, u_2, u_3 \mid \mathbf{y}_0^{\Lambda-1}, u_0) = \frac{1}{c} \prod_{i=0}^{3} W_{0,4\beta+i}^{(0)}(c_i \mid y_i) \quad (18)$$

for $\varphi = 1$, where $\mathbf{c}_0^3 = \mathbf{u}_0^3 \mathbf{B}_4 \mathbf{G}_4$ in (17) and (18).

### D. Pseudocode Description of SC Decoding

In order to extend the presented SC decoder to SCL decoding as in [9], we use a similar code structure with the main function in Alg. 1, the function recursivelyCalcP in Alg. 2 that recursively calculates the probabilities according to (13)–(18) and the function recursivelyUpdateB in Alg. 3 to update the estimated bits in the graph. We assume that the data structures $P_{\lambda,\beta}[\mathbf{u}]$ and $B_{\lambda,\beta}[\varphi]$ and the block length $n$ and $m$ are globally available in all functions. $P_{\lambda,\beta}[\mathbf{u}]$ and $B_{\lambda,\beta}[\varphi]$ can be used as three dimensional arrays where the first dimension represents the layer $0 \leq \lambda \leq m$ and the second dimension represents the branch $0 \leq \beta < 2^{m-\lambda}$. The third dimension in $B_{\lambda,\beta}[\varphi]$ indexes the bits of a branch $\beta$

**Algorithm 2:** recursivelyCalcP($\lambda, \phi$)

**Input:** $\lambda$ and $\varphi$

1. **if** $\lambda = 2$ **then**
2.     init3BitChannels($\varphi$)
3.     **return**
4. $\psi = \lfloor (\varphi - 1)/2 \rfloor$
5. **if** $\varphi = 0$ **then**
6.     recursivelyCalcP($\lambda - 1, 0$)
7. **else if** $\varphi \mod 2 \neq 0$ **and** $\varphi \neq 1$ **and** $\varphi \neq 2^\lambda - 3$ **then**
8.     recursivelyCalcP($\lambda - 1, \psi$)
9. **for** $\beta = 0, 1, \ldots, 2^{m-\lambda} - 1$ **do**
10.     $c \leftarrow 0$
11.     **for** $(u_0, u_1, u_2) \in \{0, 1\}$ **do**
12.       $P_{\lambda,\beta}[\boldsymbol{u}_0^2] \leftarrow 0$
13.       **if** $\varphi = 0$ **then**
14.         **for** $(t_0, t_1, t_2) \in \{0,1\}$ **do**
15.           $\boldsymbol{u}_b \leftarrow (u_1 \oplus u_2, t_0, t_2)$
16.           $\boldsymbol{u}_t \leftarrow (u_0 \oplus u_1 \oplus u_2, u_2 \oplus t_0, t_1)$
17.           $P_{\lambda,\beta}[\boldsymbol{u}_0^2] \leftarrow P_{\lambda,\beta}[\boldsymbol{u}_0^2]+$
18.                  $P_{\lambda-1, 2\beta}[\boldsymbol{u}_t] \cdot P_{\lambda-1, 2\beta+1}[\boldsymbol{u}_b]$
19.       **else if** $\varphi = 2^\lambda - 3$ **then**
20.         $\boldsymbol{u}_b \leftarrow (B_{\lambda,\beta}[\varphi - 2] \oplus B_{\lambda,\beta}[\varphi - 1], u_0 \oplus u_1, u_2 \oplus B_{\lambda,\beta}[0])$
21.         $\boldsymbol{u}_t \leftarrow \boldsymbol{u}_b \oplus (B_{\lambda,\beta}[\varphi - 3], B_{\lambda,\beta}[\varphi - 1], u_1)$
22.         $P_{\lambda,\beta}[\boldsymbol{u}_0^2] \leftarrow P_{\lambda-1,2\beta}[\boldsymbol{u}_t] \cdot P_{\lambda-1,2\beta+1}[\boldsymbol{u}_b]$
23.       **else if** $\varphi \mod 2 = 0$ **then**
24.         **for** $t_0 = 0, 1$ **do**
25.           $\boldsymbol{u}_b \leftarrow (B_{\lambda,\beta}[\varphi - 1] \oplus u_0, u_1 \oplus u_2, t_0)$
26.           $\boldsymbol{u}_t \leftarrow \boldsymbol{u}_b \oplus (B_{\lambda,\beta}[\varphi - 2], u_0, u_2)$
27.           $P_{\lambda,\beta}[\boldsymbol{u}_0^2] \leftarrow P_{\lambda,\beta}[\boldsymbol{u}_0^2]+$
28.                  $P_{\lambda-1,2\beta}[\boldsymbol{u}_t] \cdot P_{\lambda-1,2\beta+1}[\boldsymbol{u}_b]$
29.       **else**
30.         **for** $(t_0, t_1) \in \{0,1\}$ **do**
31.           $\boldsymbol{u}_b \leftarrow (u_0 \oplus u_1, u_2 \oplus t_0, t_1)$
32.           $\boldsymbol{u}_t \leftarrow \boldsymbol{u}_b \oplus (B_{\lambda,\beta}[\varphi - 1], u_1, t_0)$
33.           $P_{\lambda,\beta}[\boldsymbol{u}_0^2] \leftarrow P_{\lambda,\beta}[\boldsymbol{u}_0^2]+$
34.                  $P_{\lambda-1,2\beta}[\boldsymbol{u}_t] \cdot P_{\lambda-1,2\beta+1}[\boldsymbol{u}_b]$
35.       $c \leftarrow c + P_\lambda[\beta][\boldsymbol{u}_0^2]$
36.     **for** $(u_0, u_1, u_2) \in \{0,1\}$ **do**  // normalization
37.       $P_{\lambda,\beta}[\boldsymbol{u}_0^2] \leftarrow P_{\lambda,\beta}[\boldsymbol{u}_0^2]/c$

---

**Algorithm 3:** recursivelyUpdateB($\lambda, \phi$)

**Input:** $\varphi$ and $\lambda$

1. **if** $\lambda = 0$ **or** $\varphi = 0$ **or** ($\varphi \mod 2 \neq 0$ **and** $\varphi \neq 2^\lambda - 1$) **then**
2.     **return**
3. $\psi \leftarrow \lfloor (\varphi - 1)/2 \rfloor$
4. **for** $\beta = 0, 1, \ldots, 2^{m-\lambda} - 1$ **do**
5.     **if** $\varphi \mod 2 = 0$ **then**
6.       $B_{\lambda-1}[2\beta][\psi] \leftarrow \bigoplus_{i=0}^{2} B_\lambda[\beta][\varphi - i]$
7.       $B_{\lambda-1}[2\beta + 1][\psi] \leftarrow \bigoplus_{i=0}^{1} B_\lambda[\beta][\varphi - i]$
8.     **else** // last bit in subcode
9.       $B_{\lambda-1}[2\beta][\psi] \leftarrow B_\lambda[\beta][0] \oplus \bigoplus_{i=0}^{1} B_\lambda[\beta][\varphi - i]$
10.       $B_{\lambda-1}[2\beta + 1][\psi] \leftarrow B_\lambda[\beta][0] \oplus B_\lambda[\beta][\varphi]$
11. recursivelyUpdateB($\lambda - 1, \psi$)

---

**Algorithm 4:** init3BitChannels

**Input:** $\varphi$

1. **for** $\beta = 0, 1, \ldots, 2^{m-2} - 1$ **do**
2.     $c \leftarrow 0$
3.     **for** $(u_0, u_1, u_2) \in \{0,1\}$ **do**
4.       **if** $\varphi = 0$ **then**
5.         $P_{2,\lambda}[\boldsymbol{u}_0^2] = 0$
6.         **for** $t_0 = 0, 1$ **do**
7.           $(c_0, c_1, c_2, c_3) \leftarrow \text{Enc}((u_0, u_1, u_2, t_0))$
8.           $P_{2,\beta}[\boldsymbol{u}_0^2] \leftarrow P_{2,\beta}[\boldsymbol{u}_0^2] + \prod_{i=0}^{3} P_{0,4\beta+i}[c_i]$
9.       **else**
10.         $(c_0, c_1, c_2, c_3) \leftarrow \text{Enc}((B_{2,\beta}[0], u_0, u_1, u_2))$
11.         $P_{2,\beta}[\boldsymbol{u}_0^2] \leftarrow \prod_{i=0}^{3} P_{0,4\beta+i}[c_i]$
12.       $c \leftarrow c + P_{2,\beta}[\boldsymbol{u}_0^2]$
13.     **if** $\varphi = 1$ **then**  // normalization
14.       **for** $(u_0, u_1, u_2) \in \{0,1\}$ **do**
15.         $P_{2,\beta}[\boldsymbol{u}_0^2] \leftarrow P_{2,\beta}[\boldsymbol{u}_0^2]/c$

---

in layer $\lambda$. The third dimension in $P_{\lambda,\beta}[\boldsymbol{u}]$ indexes the eight values that the triple $\boldsymbol{u} = (u_\varphi, u_{\varphi+1}, u_{\varphi+2})$ can take on. For implementation we can use a binary to decimal conversion to index that triple.

In the main loop of Alg. 1, we successively calculate the probabilities $W_{m,0}^{(\varphi)}$ for $0 \leq \varphi \leq n - 3$ in Line 4. The joint distribution is marginalized over the second and third bit to get the probabilities $P_{U_\varphi|Y_0^{n-1}U_0^{\varphi-1}}(0|\mathbf{y}, \hat{\mathbf{u}}_0^{\varphi-1})$ in Line 9 and a decision on the current bit is made based on this value if the bit is not frozen. After every decision, the values of the bits in the graph are updated using the function recursivelyUpdateB in Alg. 3.

After the main loop, we get the marginalized probabilities for the last two bits from the last calculated $W_{m,0}^{(n-3)}$.

In Alg. 2, we recursively calculate the 3-bit-channel distributions. In Lines 5-8, we recursively jump to the lowest layer where the 3-bit-channel distribution is updated. When layer $\lambda = 2$ is reached, we initialize the 3-bit-channels using (17) and (18) in Alg. 4. The four cases in the main loop of recursivelyCalcP correspond to (16), (14), (15) and (13) in this order.

In Alg. 3, we recursively jump to a deeper layer $\lambda - 1$ if a change of bit $\varphi$ in layer $\lambda$ influences the next layer $\lambda - 1$.

The described decoder can be further optimized, e.g., in the main loop of Alg. 1, we calculate $P_{U_0U_1U_2|Y_1^{n-1}}, P_{U_1U_2U_3|Y_1^{n-1}U_0}, \ldots$ and marginalize them to get $P_{U_0|Y_0^{n-1}}, P_{U_1|Y_0^{n-1}U_1}, \ldots$. Certainly, we can get the marginal distributions of the first three bits out of the joint distribution of the first three bits as it is done after the main loop for the last two bits. But in order to get an easy recursive description we accept the overhead.

The presented method extends to list decoding using the same methods as in [9].

## IV. BMERA CODED MODULATION

We mimic polar coded modulation (PCM) from [10], [11], but we use multilevel BMERA codes with multistage decoding instead of multilevel polar codes. We use set partitioning labeling for the bit-levels [11, Sec. II-D.3].

We design (find the set of un/frozen bits) the BMERA codes for higher-order modulation using surrogate channels for each of the bit-levels as described in [11, Sec. IV], where we replace biAWGN surrogate channels by BEC surrogate channels. Using the BEC surrogate channels, we perform density evolution for every bit-level based on the rules presented in [5, (79a)-(80p)].

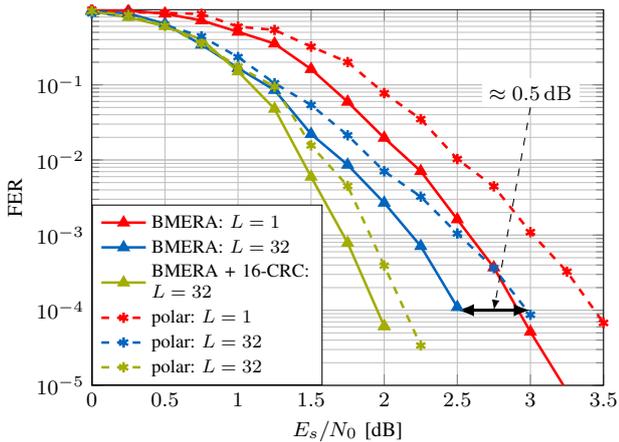

Fig. 2. SC(L) Performance of BPSK polar and BMERA codes of length $n = 1024$ with rate 0.5 over the AWGN channel.

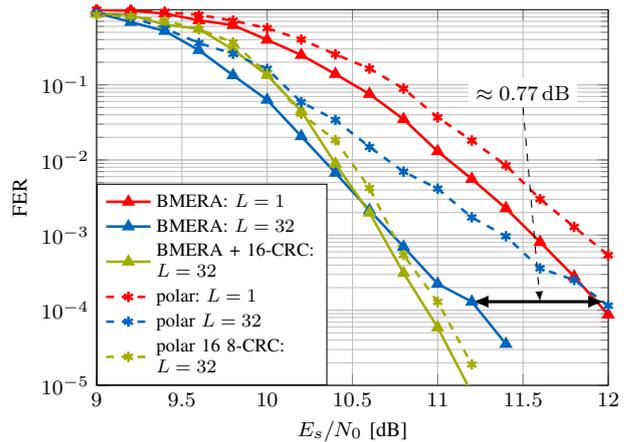

Fig. 3. SC(L) Performance of 64-QAM polar and BMERA codes of length 256 symbols per block and a spectral efficiency of 3 bits per channel use over the AWGN channel.

Demapping and decoding is done successively for every bit-level [11, Sec. II-C]. When doing list decoding for multistage decoding, it is important to keep the list across the bit-levels, i.e., the second decoder is already initialized with a list.

## V. NUMERICAL RESULTS

We provide numerical results for polar and BMERA codes with SC and SCL decoding with and without outer CRC-codes over the AWGN channel for BPSK and 64-QAM.

Fig. 2 shows the performance of BPSK polar and BMERA codes under SC(L) decoding with block length $n = 1024$ and rate 0.5. The codes are designed according to SC decoding for every simulated SNR point by using Gaussian approximation for polar codes and a BEC surrogate construction for BMERA codes [5, (79a)-(80p)]. BMERA codes clearly outperform polar codes with SC and SCL decoding. We can gain more than 0.5 dB in the presented range. Using outer CRC-codes to improve the distance spectrum of the codes [9], [12] the gain is smaller and BMERA codes are only slightly better than polar codes.

Fig. 3 shows the performance of 64-QAM polar and BMERA codes with $n = 256$ QAM symbols per block and a spectral efficiency of 3 bits per channel use over the AWGN channel. All codes are optimized for a SNR of $E_s/N_0 = 11$ dB using the methods of [11] with set partitioning label. The comparison shows a similar result as in the BPSK case. For SC and SCL decoding BMERA codes clearly outperform polar codes. After adding an outer CRC code, however, the performance of BMERA and polar codes is similar.

## VI. CONCLUSIONS

We introduced a SC decoder for BMERA (convolutional polar) codes that can be directly extended to SCL decoding. We provide a detailed pseudocode description for software implementation. BMERA codes outperform polar codes by more than 0.5 dB over the AWGN channel for BPSK and QAM. When using outer CRC codes, however, the performance of BMERA codes is only slightly better than polar codes with the same list size. Furthermore, the complexity of BMERA decoding is higher than for polar codes although both decoder complexities scale as $\mathcal{O}(Ln \log n)$. If list decoding should be avoided due to complexity reasons and to avoid list management, BMERA codes can be a promising alternative to polar codes.